# E-Learning Quality Criteria and Aspects

Reema Ajmera[1], Dinesh Kumar Dharamdasani[2]

[1]*Assistant Professor I, Deptt. Of Computer Science and Application,
JECRC University, Jaipur, India*

[2]*Assistant Professor , School of Computer and Systems Sciences,
Jaipur National University, Jaipur, India*

*Abstract-* **As IT grows the impact of new technology reflects in more or less every field. Education also gets new dimensions with the advancement in IT sector. Nowadays education is not limited to books and black boards only it gets a new way i.e. electronic media. Although with e-learning, the education having broader phenomena, yet it is in budding stage. Quality is a crucial issue for education as well as e-learning. It is required to serve qualitative and standardization education. Quality cannot be expressed and set by a simple definition, since in itself quality is a very abstract notion. The specified context and the perspectives of users need to be taken into account when defining quality in e-learning. It is also essential to classify suitable criteria to address quality.**

## I. INTRODUCTION

Electronic Learning (e-learning), is education based on modern methods of communication including the computer and its networks, various audio-visual materials, search engines, electronic libraries, and websites, whether accomplished in the classroom or at a distance. Generally speaking, this type of education is delivered through the medium of the World Wide Web where the educational institution makes its programs and materials available on a special website in such a manner that students are able to make use of them and interact with them with ease through closed or shared, networks, or the Internet, and through use of e-mail and online discussion groups. A number of other terms are also used to describe this mode of teaching and learning. They include online learning virtual learning distributed learning network and web-based learning. Fundamentally, they all refer to educational processes that utilize information and communications technology to mediate asynchronous as well as synchronous learning and teaching activities.

The definition of e-learning centers on its being a learning method and a technique for the presentation of academic curricula via the Internet or any other electronic media inclusive of multimedia, compact discs, satellites, or other new education technologies. The two parties participating in the educational process interact through these media to achieve specific educational objectives.

During the 1980s, the compact disc (CD) began to be used in education, but the fact that it lacked the quality of interaction between the student, the material and the teacher was an important flaw in the opinion of a number of educators.

This problem was only resolved with the appearance of the Internet which justified the adoption of e-learning because it fulfilled the condition of immediacy or simultaneity.

There are two main types of e-learning: asynchronous and synchronous, depending on the interaction between learner and teacher Synchronous e-learning environments require tutors and learners or the online classmates to be online at the same time, where live interactions take place between them. However, where students are logging into and using the system independently of other students and staff members are synchronous e-learning environment.[5]

E-learning assists in the transformation of the educational process from the stage of learning by rote to one characterized by creativity, interaction and the development of skills. The student, in e-learning, is able to access educational materials at any time and from any place, thereby transforming the concepts of the educational process and learning to go beyond the limits imposed by traditional classrooms into a rich environment in which there are numerous sources of learning. Sources of programs of e-learning include experts in the field, ministries, corporations and other organizations concerned with the dissemination of technical applications in education. Programs are offered by way of closed or shared networks, as well as over the Internet, and e-mail and discussion groups are among the techniques and mechanisms employed in e-learning

## II. MODALS OF E-LEARNING

These various types or modalities of e-learning activity are represented in fig.1 [4]

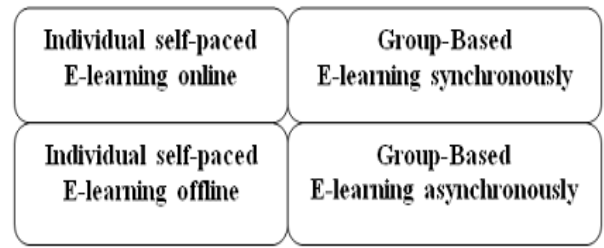

**Individualized self-paced e-learning online** refers to situations where an individual learner is accessing learning resources such as a database or course content online via an Intranet or the Internet.

A typical example of this is a learner studying alone or conducting some research on the Internet or a local network.

**Individualized self-paced e-learning offline** refers to situations where an individual learner is using learning resources such as a database or a computer-assisted learning package offline (i.e., while not connected to an Intranet or the Internet). An example of this is a learner working alone off a hard drive, a CD or DVD.

**Group-based e-learning synchronously** refers to situations where groups of learners are working together in real time via an Intranet or the Internet. It may include text-based conferencing, and one or two-way audio and videoconferencing. Examples of this include learners engaged in a real-time chat or an audio-videoconference.

**Group-based e-learning asynchronously** refers to situations where groups of learners are working over an Intranet or the Internet where exchanges among participants occur with a time delay (i.e.not in real time). Typical examples of this kind of activity include on-line discussions via electronic mailing lists and text-based conferencing within learning managements systems

### III. METHOD AND PRACTICES OF E-LEARNING

Learner and learning-centered educational processes are defining characteristics of situated learning environments. The concept of situated learning is grounded in the principles of constructivist learning theory (Wilson, 1996). It is based on the belief that learning is most efficient and effective when it takes place within the context of realistic educational settings which are either real or contrived (see Brown, Collins & Duguid, 1989).

*Scenario based learning*

A good learning scenario will reflect a common occurrence from the relevant field [2].

A typically good learning scenario will sound like a story or a narrative of a common occurrence. It will have a context, a plot, characters and other related parameters. It usually involves a precipitating event which places the learner or a group of learners in a role, or roles that will require them to deal with the situation or problems caused by the event. The roles that learners might be asked to assume are those that they are likely to play in real life as they enter the workforce. Attached to these roles, will be goals that learners will be required to achieve. In order to achieve these goals they will be assigned numerous tasks and activities, some of which may require them to collaborate with their peers and other relevant groups, if these are part of the intended learning outcomes of their subject. While these activities essentially serve as learning enhancement exercises, a selection of them could be made assessable and given a mark which would contribute to the student's final grade in the subject.

*Problem based learning*

Problem-based learning begins with the introduction of an ill-structured problem on which all learning is centered. Problem-based learning (PBL) is an approach that challenges students to learn through engagement in a real problem. It is a format that simultaneously develops both problem solving strategies and disciplinary knowledge bases and skills by placing students in the active role of problem-solvers confronted with an ill-structured situation that simulates the kind of problems they are likely to face as future managers in complex organizations.

Problem-based learning is student-centered. PBL makes a fundamental shift--from a focus on teaching to a focus on learning. The process is aimed at using the power of authentic problem solving to engage students and enhance their learning and motivation. There are several unique aspects that define the PBL approach:

*Case based learning*

A case-based approach engages students in discussion of specific scenarios that resemble or typically are real-world examples. This method is learner-centered with intense interaction between participants as they build their knowledge and work together as a group to examine the case. The instructor's role is that of a facilitator while the students collaboratively analyze and address problems and resolve questions that have no single right answer.

*Learning by designing*

In learning by designing, the design task affords the essential anchor and scaffold for all learning and teaching activities (Newstetter, 2000). In this learning design students are required to engage in a learning activity which comprises conceptualizing and building something. This is a common learning and teaching activity in the study of architecture, and engineering sciences. As in goal-based learning, in the case of learning by designing, the goal is made very clear to the students. How the students chose to pursue that goal and achieve the targeted learning outcomes is left to their imagination and creativity (see Naidu, Anderson & Riddle, 2000).

*Role-Play Based learning*

In role-play-based learning, the role-play provides the anchor and scaffold for all learning and teaching activities (see Ip & Linser, 1999; Linser, Naidu & Ip, 1999). Role-play is widely used as a valuable learning and teaching strategy in social sciences and humanities subjects where very complex processes are prevalent. This learning design comprises the playing out of identified roles by learners which is followed with reflection upon the activity and its analysis in order to focus attention on the expected learning outcomes for the study.

### IV. E-LEARNING.QUALITY ATTRIBUTES

When we talk about quality in e-learning, we assume an implicit consensus about the term 'quality'. In fact, however, 'quality' means very different things to most e-learning providers. Harvey and Green have suggested the following set of categories:

(a) exceptionality,
(b) perfection or consistency,
(c) fitness for purpose,

(d) adequate return,
(e) transformation

The last perception of quality, transformation, is the most relevant to the pedagogical process. It describes the increase in competence or ability as a result of the learning process as transformation. In order to make these categories manageable for respondents, they were operationalised as follows in the study:

A key question to be clarified by the study in this area was the picture of quality competence in the individual countries or regions. The study shows that the individual dimensions of quality are distributed very unevenly across the regions when it comes to dealing with quality strategies. The investigation focused on two constructs in particular:

(a) Knowledge of quality, which ascertains the awareness and familiarity with the topic of those who develop, use or learn from e-learning;

(b) experience of quality, which looks at length of experience of putting quality development measures into practice.
Consistency, interoperability, reusability, scalability are [1].
Quality of e-learning stressed different elements/criteria/positions among the possible ones:

Industry providing e-learning : conformance, interoperability, standardization, provision of scalable integrated learning services, product oriented process;

Industry seen as content provider: competence and expertise of the producer of the educational material, content oriented and production quality process;

Government at EU, national and local level: control on content and curriculum, control on equal opportunities, equal access, , protecting learners as customers, improving efficiency and effectiveness of learning processes;

School education: customer satisfaction, curricula integration, educational value and use of learning services, user - friendliness and usability of resources;

Higher Education: Material/ content is scientifically state-of-the-art and maintained up-to-date, prestige and recognition of the author, accreditation;

Initial vocational education and training: support to contextualization, quality of the product, clearly explicit pedagogical design principles appropriate to learner type, needs and context, high level of interactivity;

Informal learning: Accessibility by different target groups in particular the ones have been excluded before, low-cost, support to individual path, availability of support mechanism, that help people overcome an obstacle that might have prevent them from engaging in formal learning;

Continuous professional development: content of the programme and the quality of resources, accreditation system for centres to deliver their qualification programs, relevance to work processes and working contexts.

**SCORM**
SCORM is an international standard for tracking E-Learning activities.
Sharable Content Object Reference Model (SCORM) is a collection of standards and specifications for web-based e-learning. It defines communications between client side content and a host system (called "the run-time environment"), which is commonly supported by a learning management system. SCORM also defines how content may be packaged into a transferable ZIP file called Package Interchange Format. SCORM is a specification of the Advanced Distributed Learning (ADL) Initiative from the Office of the United States Secretary of Defense.

SCORM 2004 introduced a complex idea called sequencing, which is a set of rules that specifies the order in which a learner may experience content objects. In simple terms, they constrain a learner to a fixed set of paths through the training material, permit the learner to "bookmark" their progress when taking breaks, and assure the acceptability of test scores achieved by the learner. The standard uses XML, and it is based on the results of work done by AICC, IMS Global, IEEE, and Ariadne. [3]

## V. E-LEARNING QUALITY MODEL

Quality is an evaluation of excellence and quality can be viewed and considered by different aspects so it is important to set standards for e-learning quality, this is a difficult and complex issue because there is no formal definition of information quality, as quality is dependent on the criteria applied to it. Furthermore, it is dependent on the targets, the environment and from which viewpoint.[5]

ELQM is made up of four quality aspects which we consider crucial when assessing quality in e-learning that are cohesive on other factors:

Delivery- It can be observed that content is the most critical component of learning through the internet. The most important aspect of delivering the content is the quality of the content for the client. If the content delivered is good than only the student can be assessed properly. By good content we can say that a particular student enrolled for a particular course must get the contents related to his course and in proper manner. Course material must be designed in such a way so that it can provide flexibility and adaptability to the student. This can be done by including the diagrams and pictures to illustrate the whole scenario. Student enrolled should get the suitable resources required for the successful completion of the course.

Assessment: It is a very important part of overall course curriculum in which the student is assessed for his understanding during the total duration of the course. Student must be assessed on every aspect of evaluation so that the quality of the course can be judged. One time assessment of student on a particular evaluation method will not be able to judge as the continuous evaluation is needed if a student has to be judged in a satisfactory manner. These assessments can be made on different techniques like by putting a student in a real time situation and ask him to give the solution of a particular problem. Student can also be judged by means of questionnaires.

Trainer/Assessor Competence- Selection of a trainer or assessor plays a vital role in the successful running of any training or course. While selecting a trainer for any course it is important to know about the capabilities of the trainer in the respective field. Capabilities can be related to the qualification and the experience of the trainer. A trainer must be selected keeping in view the strength and weaknesses of the students for which the trainer has to be selected. Trainer must be a person who can deliver his contents to the students in a effective manner. He must check regularly on a regular basis the progress of a student and change his delivery method if needed by the student. Trainer is also responsible for making the sessions interactive by the cooperation of the students he is delivering the contents. A effective trainer is a person who is responsible for the success and failure of any course or training.

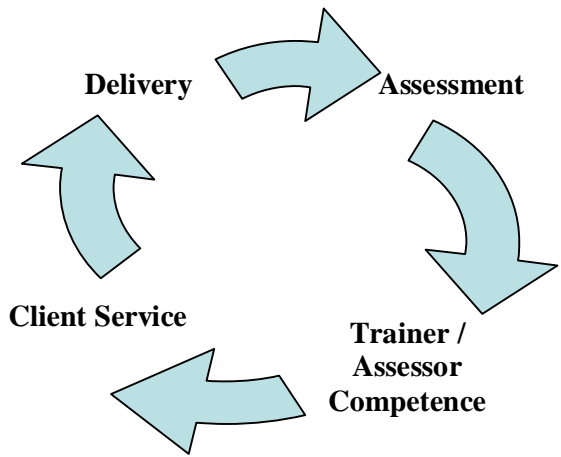

Client Service: Nowadays education is not limited to the boundaries of library or classroom it has been totally transformed by the intervention of internet or electronic media. Although with e-learning, the education having broader phenomena, yet it is in crucial stage. Ensuring good quality is a very important issue for education in the electronic media. Before enrolling for any course or training in the online mode student wants to be aware about the vision and institutional leadership qualities of the organization providing the courses in the electronic mode. In the online mode a student want a clear structure of the course he is interested in as there is everything in the virtual environment and nothing is in front of him. Student want to know about the working interface and the support which he will get after enrolling in a particular course. Student must be convinced regarding each step involved from beginning to end so that there is no ambiguity in understanding the whole process.

The aspects above are not numbered in order of importance, but there is a rough sequence from the smallest elements of teaching/learning processes to an organizational, systemic and holistic view. This in fact also reflects the two different and complementary sources of information we have used in this study: those with an organizational perspective and those with a research perspective.

## VI. FUTURE ASPECT

The fundamental obstacle to the growth of e-learning is lack of access to the necessary technology infrastructure, for without it there can be no e-learning. Poor or insufficient technology infrastructure is just as bad, as it can lead to unsavory experiences that can cause more damage than good to teachers, students and the learning experience. While the costs of the hardware and software are falling, often there are other costs that have often not been factored into the deployment of e-learning ventures. The most important of these include the costs of infrastructure support and its maintenance, and appropriate training of staff to enable them to make the most of the technology [2].